\documentstyle[prd,preprint,tighten,aps,eqsecnum,%
amssymb,newlfont,graphicx]{revtex}
\begin{document}
\draft
\preprint{
\begin{tabular}{r}
IASSNS-AST 97/68\\
JHU-TIPAC 97020\\
KIAS-P97011\\
DFTT 72/97\\
hep-ph/9711400
\end{tabular}
}
\title{Neutrino oscillation constraints on neutrinoless double beta decay}
\author{S. M. Bilenky}
\address{Joint Institute for Nuclear Research, Dubna, Russia, and\\
Institute for Advanced Study, Princeton, N.J. 08540}
\author{C. Giunti}
\address{INFN, Sezione di Torino, and Dipartimento di Fisica Teorica,
Universit\`a di Torino,\\
Via P. Giuria 1, I--10125 Torino, Italy}
\author{C. W. Kim}
\address{Department of Physics $\&$ Astronomy,
The Johns Hopkins University,\\
Baltimore, MD 21218, USA, and\\
School of Physics,
Korea Institute for Advanced Study,
Seoul 130-012, Korea}
\author{M. Monteno}
\address{INFN, Sezione di Torino, and Dipartimento di Fisica Sperimentale,
Universit\`a di Torino,\\
Via P. Giuria 1, I--10125 Torino, Italy}
\date{November 20, 1997}
\maketitle
\begin{abstract}
We have studied the constraints imposed by
the results of neutrino oscillation experiments
on the effective Majorana mass
$|\langle{m}\rangle|$
that characterizes the contribution of Majorana neutrino masses
to the matrix element of
neutrinoless double-beta decay.
We have shown that
in a general scheme
with three Majorana neutrinos and
a hierarchy of neutrino masses
(which can be explained by the see-saw mechanism), 
the results of neutrino oscillation 
experiments imply rather strong constraints on the parameter 
$|\langle{m}\rangle|$.
From the results of the first reactor long-baseline experiment CHOOZ
and the Bugey experiment
it follows that
$ |\langle{m}\rangle| \lesssim 3 \times 10^{-2} \, \mathrm{eV} $
if
$ \Delta{m}^2 \lesssim 2 \mathrm{eV}^2 $
($\Delta{m}^2$
is the largest mass-squared difference).
Hence,
we conclude that
the observation of neutrinoless double-beta decay
with a probability that corresponds to
$ |\langle{m}\rangle| \gtrsim 10^{-1} \, \mathrm{eV} $
would be a signal for a non-hierarchical neutrino mass 
spectrum and/or non-standard mechanisms of lepton number violation.
\end{abstract}


\section{Introduction}
\label{Introduction}

The investigation of the fundamental properties of neutrinos
(neutrino masses and neutrino mixing,
the nature of massive neutrinos (Dirac or Majorana?),
neutrino magnetic moments, etc.)
is the most important problem of today's neutrino physics.
This investigation is one of the major directions
of search for 
physics beyond the Standard Model.

At present,
there are three experimental 
indications in favor of neutrino oscillations.
The first indication comes from 
solar neutrino experiments
(Homestake \cite{Homestake},
Kamiokande \cite{Kam-sol},
GALLEX \cite{GALLEX},
SAGE \cite{SAGE}
and
Super-Kamiokande \cite{SK-sol}).
The second indication
was found
in the
Kamiokande \cite{Kam-atm},
IMB \cite{IMB},
Soudan \cite{Soudan}
and
Super-Kamiokande \cite{SK-atm}
atmospheric neutrino experiments.
The third indication in favor of neutrino oscillations
was obtained by the LSND
collaboration
\cite{LSND95,LSND96}.
On the other hand,
in many short-baseline (SBL) reactor and accelerator experiments 
(see the reviews in Refs.\cite{Boehm-Vannucci}) and in the recent 
long-baseline (LBL) reactor experiment CHOOZ \cite{CHOOZ97}
no indications in favor of neutrino
oscillations were found.

Neutrino oscillation experiments
cannot provide an answer to the question:
what type of particles are massive neutrinos,
Dirac or Majorana? 
(see Ref.\cite{BHP80}).
The answer to this question,
that
is of fundamental importance,
could be obtained
from experiments
on the investigation of  processes in which
the total lepton number 
$ L=L_{e}+L_{\mu}+L_{\tau} $
is not conserved.
The classical process of this type is
neutrinoless double-$\beta$ decay
($(\beta\beta)_{0\nu}$)
\begin{equation}
(A,Z)
\to
(A,Z+2)
+
e^{-}
+
e^{-}
\,.
\label{01}
\end{equation}

The neutrinoless double-$\beta$ decay
of different nuclei
has been searched for in many experiments
(see, for example, Ref.\cite{BBreviews}).
No positive signal was found
up to now.
The most stringent limits on the half-lives
for $(\beta\beta)_{0\nu}$ decay were found in
$^{76}$Ge and $^{136}$Xe experiments.
In the experiments
of the Heidelberg-Moscow
\cite{Heidelberg-Moscow}
and
Caltech-Neuchatel-PSI
\cite{Caltech-Neuchatel-PSI}
collaborations
it was found that
\begin{eqnarray}
T_{1/2}(^{76}\mathrm{Ge})
>
7.4 \times 10^{24} \, \mathrm{y}
& \qquad \mbox{(90\% CL)} \qquad &
\mbox{Heidelberg-Moscow}
\,,
\label{02}
\\
T_{1/2}(^{136}\mathrm{Xe})
>
4.2 \times 10^{23} \, \mathrm{y}
& \qquad \mbox{(90\% CL)} \qquad &
\mbox{Caltech-Neuchatel-PSI}
\,.
\label{03}
\end{eqnarray}

The standard mechanism of  $(\beta\beta)_{0\nu}$ decay
is the mechanism of mixing of neutrinos with Majorana masses.
In accordance with the hypothesis of neutrino
mixing
(see Refs.\cite{BP87,CWKim93,Mohapatra-Pal}),
the left-handed flavor neutrino fields
$ \nu_{{\ell}L} $
are combination of fields of neutrinos with definite masses:
\begin{equation}
\nu_{{\ell}L}
=
\sum_{i}
U_{{\ell}i}
\,
\nu_{iL}
\qquad \qquad
(\ell=e,\mu,\tau)
\,,
\label{05}
\end{equation}
where $ \nu_{i} $ is the field of neutrinos with mass $m_i$ and
$U$ is the unitary mixing matrix.
If massive neutrinos are Majorana particles,
the fields  $ \nu_{i} $ satisfy the Majorana condition 
$ \nu_{i} = \nu_{i}^{c} \equiv C \overline{\nu}_{i}^{T} $
($C$ is the charge conjugation matrix),
 the total lepton number is not conserved and
$(\beta\beta)_{0\nu}$ decay is possible.
In the framework of neutrino mixing 
the process (\ref{01}) is
a process of the second order in the CC weak interaction
hamiltonian
\begin{equation}
\mathcal{H}_{I}
=
\frac{ G_{F} }{ \sqrt{2} }
\,
2
\sum_{\ell=e,\mu,\tau}
\overline{\ell}_{L}
\,
\gamma^{\alpha}
\,
\nu_{{\ell}L}
\, \,
j_{\alpha}^{\mathrm{CC}}
+
\mathrm{h.c.}
\,,
\label{04}
\end{equation}
with a virtual neutrino.
In (\ref{04})
$G_{F}$ is the Fermi constant and
$j_{\alpha}^{\mathrm{CC}}$ is the standard hadronic charged current.
The matrix element
of $(\beta\beta)_{0\nu}$ decay is proportional to the effective Majorana
neutrino mass
(see, for example, Refs.\cite{BP87,CWKim93,Mohapatra-Pal})
\begin{equation}
\langle{m}\rangle
=
\sum_{i}
U_{ei}^2
\,
m_{i}
\,.
\label{07}
\end{equation}

The negative results of the experiments searching for
$(\beta\beta)_{0\nu}$ decay
imply upper bounds for the
the parameter
$|\langle{m}\rangle|$.
The numerical values of the upper bounds
depend on the model that is used for the calculation
of the nuclear matrix elements.
From
the results of the
$^{76}$Ge
and
$^{136}$Xe
experiments 
the following upper bounds were obtained:
\begin{eqnarray}
\null & \null & \null
|\langle{m}\rangle|
<
( 0.6 - 1.1 ) \, \mathrm{eV}
\qquad
\qquad
\mbox{($^{76}$Ge \cite{Heidelberg-Moscow,Faessler})}
\,,
\label{081}
\\
\null & \null & \null
|\langle{m}\rangle|
<
( 2.3 - 2.7 ) \, \mathrm{eV}
\qquad
\qquad
\mbox{($^{136}$Xe \cite{Caltech-Neuchatel-PSI})}
\,.
\label{09}
\end{eqnarray}
A significant progress in search of
neutrinoless double-$\beta$ decay
is expected in the future.
Several collaborations are planning to reach
a sensitivity of
$ 0.1 - 0.3 \, \mathrm{eV} $
for
$|\langle{m}\rangle|$
\cite{Heidelberg-Moscow,futureBB}.

Contributions to the matrix element of
$(\beta\beta)_{0\nu}$ decay 
of different non-standard mechanisms
for violation of the lepton number
(right-handed currents \cite{Mohapatra-Pal,Mohapatra95,HKP},
supersymmetry with violation of R-parity \cite{BM95,Mohapatra95,HKK-susy,FKSS},
and others \cite{HKK-others,PCSW})
have recently been considered in the literature.
At present,
it is not possible to distinguish different mechanisms.
It is obvious that it is important to
obtain independent information about
the contribution to the matrix element of
$(\beta\beta)_{0\nu}$ decay
of Majorana neutrino masses and mixing, 
given by the effective Majorana neutrino mass
$|\langle{m}\rangle|$.

In this paper,
we will show that the existing neutrino oscillation data imply rather
strong constraints on the effective Majorana mass
$|\langle{m}\rangle|$
under the general assumption of a neutrino mass hierarchy.
The first estimates of the parameter
$|\langle{m}\rangle|$
obtained from the data of SBL reactor experiments were given in
Ref.\cite{PS94}
and
a more detailed analysis,
including the results of the
Krasnoyarsk \cite{Krasnoyarsk94}
and
Bugey \cite{Bugey95}
experiments 
and the first results of the LSND experiment \cite{LSND95}
was presented in Ref.\cite{BBGK}.
Since these analyses have been carried out,
new results of the LSND experiment
have been published \cite{LSND96}
and
the results of the first LBL reactor experiment CHOOZ
appeared \cite{CHOOZ97}.
We will use all these data
and the results of the Kamiokande \cite{Kam-atm}
and Super-Kamiokande \cite{SK-atm}
atmospheric neutrino experiments
in order to obtain new bounds on the effective
Majorana mass
$|\langle{m}\rangle|$.
In Sections
\ref{Constraints from reactor neutrino experiments and the LSND experiment}
and
\ref{Constraints from atmospheric neutrino experiments}
we will see that these data imply rather strong 
constraints on this parameter.
In Section \ref{Non-hierarchical neutrino mass spectra}
we present some remarks on non-hierarchical
neutrino mass spectra.

\section{Three neutrinos with a mass hierarchy}
\label{Three neutrinos with a mass hierarchy}

The results of the LEP experiments
on the measurement of the invisible 
width of the $Z$ boson
imply that only three flavor neutrinos exist in nature
(see Ref.\cite{RPP}).
The number of light massive Majorana neutrinos 
is equal to three in the case of 
a left-handed
Majorana mass term and can be more than three
in the general
case of a Dirac and Majorana mass term
(see, for example, Refs.\cite{BP87,CWKim93,Mohapatra-Pal}).
Let us notice that the result of LEP
measurements does not exclude this last possibility.
If the number of light massive Majorana neutrinos is more than three,
sterile neutrinos must exist.
The sterile fields do not enter in the standard neutral current
and their effect is not seen in LEP experiments.

We will consider here the simplest case
of three light Majorana neutrinos\footnote{Some
remarks about the case of four neutrinos are presented
in Section \ref{Non-hierarchical neutrino mass spectra}.}.
As is well known,
a general characteristic
feature of the mass spectra of
leptons and quarks
is the hierarchy of the masses of the particles
of different generations.
What about neutrinos?
Different possibilities for
the mass spectrum
of three neutrinos were considered in the
literature (see
\cite{BBGK,three-hierarchy,three-inverted,BGKP}).
We assume that
the neutrino masses
$m_1$, $m_2$, $m_3$,
as in the case of the masses of quarks and leptons,
satisfy the hierarchy\footnote{Another
possible mass spectrum of three neutrino is discussed
in Section \ref{Non-hierarchical neutrino mass spectra}.}
\begin{equation}
m_1 \ll m_2 \ll m_3
\,.
\label{10}
\end{equation}
Such a spectrum corresponds to the 
see-saw mechanism for neutrino mass generation
\cite{see-saw}
which is the only known mechanism that
explains naturally the smallness of neutrino masses
with respect to the masses of other fundamental fermions.
We do not assume,
however,
any specific (quadratic or linear) see-saw relation between neutrino masses.
We will use only the results
of neutrino oscillation experiments
in the general framework of a hierarchy (\ref{10})
of neutrino masses.

In all solar neutrino experiments
(Homestake \cite{Homestake},
Kamiokande \cite{Kam-sol},
GALLEX \cite{GALLEX},
SAGE \cite{SAGE}
and
Super-Kamiokande \cite{SK-sol})
the detected event rates are significantly
smaller than the event rates predicted by the existing 
Standard Solar Models (SSM)
\cite{SSM}.
Moreover,
a phenomenological analysis
of the data of the different solar neutrino experiments,
in which the values of the neutrino fluxes
predicted by the SSM are not used,
strongly suggest that
the solar neutrino problem is real
\cite{phenomenological}.
In order to take into account
the results of solar
neutrino experiments
in the framework of a
hierarchy of neutrino masses,
it is necessary to assume 
that
$
\Delta{m}^2_{21}
\equiv
m_2^2 - m_1^2
$
is relevant for the suppression
of the flux of solar $\nu_e$'s.
In this case, 
the results of the solar neutrino experiments
and the predictions of the SSM
can be reconciled if
\begin{equation}
\Delta{m}^{2}_{21}
\sim
( 0.3 - 1.2 ) \times 10^{-5}\, \mathrm{eV}^2
\qquad
\mbox{or}
\qquad
\Delta{m}^{2}_{21} \sim 10^{-10}\, \mathrm{eV}^2
\,,
\label{11}
\end{equation}
in the case of
MSW resonant transitions
\cite{SOLMSW}
and
just-so vacuum oscillations
\cite{SOLVAC},
respectively.

Under the assumption of a neutrino mass hierarchy,
the effective Majorana mass
$|\langle{m}\rangle|$
is given by \cite{PS94}
\begin{equation}
|\langle{m}\rangle|
\simeq
|U_{e3}|^2 \, m_3
\simeq
|U_{e3}|^2 \, \sqrt{ \Delta{m}^{2} }
\,,
\label{13}
\end{equation}
with
$
\Delta{m}^{2}
\equiv
m_3^2 - m_1^2
$.

\section{Constraints from reactor neutrino experiments and the LSND experiment}
\label{Constraints from reactor neutrino experiments and the LSND experiment}

In order to obtain
information on
$|U_{e3}|^2$
and the effective Majorana mass
$|\langle{m}\rangle|$
from the results of
reactor oscillation experiments,
we will follow the method presented in
Ref.\cite{BBGK}
(see also Ref.\cite{PS94}).

In the case of a small 
$ \Delta{m}^{2}_{21} $
and a neutrino mass hierarchy, 
the probability
of
the transitions
$ \nu_{\ell} \to \nu_{\ell'} $
of terrestrial neutrinos
is given by
\begin{equation}
P_{\nu_{\ell}\to\nu_{\ell'}}
=
\left|
\delta_{{\ell'}{\ell}}
+
U_{{\ell'}3}
\,
U_{{\ell}3}^{*}
\left(
e^{ - i \frac{ \Delta{m}^{2} L }{ 2 p } }
-
1
\right)
\right|^2
\,.
\label{14}
\end{equation}
Here $L$ is the distance between
the neutrino source and the detector
and
$p$ is the neutrino momentum.
In Eq.(\ref{14}) we used the unitarity of the mixing matrix and
we took into account the fact
that for the distances and energies
of neutrinos in terrestrial experiments
$ \Delta{m}^{2}_{21} L / 2 p \ll 1 $.
For the $\nu_{\ell}$
($\bar\nu_{\ell}$)
survival probability,
from Eq.(\ref{14}) we have
(see Ref.\cite{BBGK})
\begin{equation}
P_{\nu_{\ell}\to\nu_{\ell}}
=
P_{\bar\nu_{\ell}\to\bar\nu_{\ell}}
=
1
-
\frac{ 1 }{ 2 }
\,
B_{{\ell};{\ell}}
\left(
1
-
\cos
\frac{ \Delta{m}^{2} L }{ 2 p }
\right)
\,,
\label{17}
\end{equation}
where the oscillation amplitudes
$B_{{\ell};{\ell}}$
are given by
\begin{equation}
B_{{\ell};{\ell}}
=
4 \, |U_{{\ell}3}|^2
\left(
1
-
|U_{{\ell}3}|^2
\right)
\,.
\label{19}
\end{equation}

Several SBL oscillation experiments with 
reactor $\bar\nu_e$'s
have been performed
in the last years
(see the review in Ref.\cite{Boehm-Vannucci}).
No indications in favor of neutrino oscillations
were found in these experiments.
In our analysis,
we will use the exclusion plot of the 
Bugey \cite{Bugey95}
experiment
and
the recently published
results of the first LBL neutrino reactor
experiment CHOOZ \cite{CHOOZ97}
(the inclusion of the results of the
Krasnoyarsk \cite{Krasnoyarsk94} experiment
in the analysis
does not add any new constraint).
These experimental results
provide the most stringent
limits for the neutrino oscillation parameter
$B_{e;e}$.

We will consider
the square of the largest neutrino mass
$ m_3^2 \simeq \Delta{m}^{2} $
as a
parameter 
and we will consider values of this parameter
in the wide range of sensitivity of SBL and LBL
reactor neutrino experiments
\begin{equation}
10^{-3} \, \mathrm{eV}^2
\leq
\Delta{m}^{2}
\leq
10^{3} \, \mathrm{eV}^2
\,.
\label{wide}
\end{equation}

From the 90\% CL exclusion plots 
of reactor neutrino experiments,
at any fixed value of
$\Delta{m}^{2}$
in the range (\ref{wide}),
the amplitude $B_{{e};{e}}$ of 
$ \bar\nu_{e} \to \bar\nu_{e} $
transitions
is bounded by
\begin{equation}
B_{{e};{e}}
\leq
B_{{e};{e}}^{0}
\,.
\label{21}
\end{equation}
From Eqs.(\ref{19}) and (\ref{21}),
it follows that
$|U_{e3}|^2$
must satisfy one of the two inequalities:
\begin{eqnarray}
\null & \null & \null
|U_{e3}|^2
\leq
a_{e}^{0}
\,,
\label{22}
\\
\mbox{or}
\qquad
\null & \null & \null
\nonumber
\\
\null & \null & \null
|U_{e3}|^2
\geq
1 - a_{e}^{0}
\,,
\label{23}
\end{eqnarray}
where
\begin{equation}
a_{e}^{0}
\equiv
\frac{ 1 }{ 2 }
\left(
1
-
\sqrt{
1
-
B_{{e};{e}}^{0}
}
\right)
\,.
\label{24}
\end{equation}
In Fig.\ref{fig1},
we have plotted the values of
the parameter
$a_{e}^{0}$
obtained from
the 90\% CL
exclusion plots of the Bugey and CHOOZ experiments.
Figure \ref{fig1}
shows that
$a_{e}^{0}$
is small for
$\Delta{m}^{2}$
in the range (\ref{wide}).
Thus,
the results of the reactor oscillation experiments 
imply that
$|U_{e3}|^2$
can only be small
or large (close to one).

The results of the
solar neutrino experiments exclude the possibility
of a large value of $|U_{e3}|^2$.
The argument goes as follows.
The averaged probability
$ P_{\nu_e\to\nu_e}^{\mathrm{sun}}(E) $
for solar $\nu_e$'s
to survive,
in the case of a neutrino mass hierarchy with
$ \Delta{m}^{2}_{21} $
relevant for the oscillations of solar neutrinos,
is given by
(see Ref.\cite{SS92})
\begin{equation}
P_{\nu_e\to\nu_e}^{\mathrm{sun}}(E)
=
\left(
1
-
|U_{e3}|^2
\right)^2
P_{\nu_e\to\nu_e}^{(1,2)}(E)
+
|U_{e3}|^4
\,,
\label{25}
\end{equation}
where
$ P_{\nu_e\to\nu_e}^{(1,2)}(E) $
is the $\nu_{e}$ survival probability 
due to the mixing of $\nu_1$ and $\nu_2$
and
$E$ is the neutrino energy.
If
$\left| U_{e3} \right|^2$
satisfies the inequality (\ref{23}),
we have
$
P_{\nu_e\to\nu_e}^{\mathrm{sun}}(E)
\geq
( 1 - a_{e}^{0} )^2
\equiv
(P_{\nu_e\to\nu_e}^{\mathrm{sun}})_{\mathrm{min}}
$.
In
Fig.\ref{fig2}
we have plotted the values of
$ (P_{\nu_e\to\nu_e}^{\mathrm{sun}})_{\mathrm{min}} $
obtained from the
exclusion plots of the Bugey and CHOOZ experiments.
It can be seen that
$ (P_{\nu_e\to\nu_e}^{\mathrm{sun}})_{\mathrm{min}} \simeq 0.9 $
for
$ \Delta{m}^{2} \gtrsim 2 \times 10^{-3} \, \mathrm{eV}^2 $.
Furthermore,
Eq.(\ref{25}) implies that
the maximal variation of
$P_{\nu_e\to\nu_e}^{\mathrm{sun}}(E)$
as a function of neutrino energy
is given by
$ ( 1 - |U_{e3}|^2 )^2 $.
If
$ |U_{e3}|^2$
satisfies the inequality (\ref{23}),
we have
$ ( 1 - |U_{e3}|^2 )^2 \leq (a_{e}^{0})^2 $,
which is a very small quantity
(from Fig.\ref{fig1}
one can see that
$ ( 1 - |U_{e3}|^2 )^2 \lesssim 9 \times 10^{-2} $
for
$\Delta{m}^{2}$ in the range (\ref{wide})
and
$ ( 1 - |U_{e3}|^2 )^2 \lesssim 4 \times 10^{-3} $
for
$ \Delta{m}^{2} \gtrsim 2 \times 10^{-3} \, \mathrm{eV}^2 $).
Thus,
if $ |U_{e3}|^2$ is large,
$ P_{\nu_e\to\nu_e}^{\mathrm{sun}}(E) $
is practically constant.
The
large lower bound for
the survival probability
$ P_{\nu_e\to\nu_e}^{\mathrm{sun}} $
and its practical independence of the neutrino energy
are not compatible with the data of the solar neutrino experiments
(see Refs.\cite{KP97,CMMV97}).
Therefore,
from the results of solar 
and reactor neutrino experiments
we can conclude that
$|U_{e3}|^2$
is small and satisfies the
inequality (\ref{22}).

The limit (\ref{22}) for
$|U_{e3}|^2$
implies
the following upper bound for the effective Majorana mass
$|\langle{m}\rangle|$:
\begin{equation}
|\langle{m}\rangle|
\leq
a_{e}^{0}
\,
\sqrt{ \Delta{m}^{2} }
\,.
\label{26}
\end{equation}
The upper bounds
obtained with Eq.(\ref{26}) from the
90\% CL exclusion plots
of the
Bugey \cite{Bugey95}
and
CHOOZ \cite{CHOOZ97}
experiments
for
$
10^{-4} \, \mathrm{eV}^2
\leq
\Delta{m}^{2}
\leq
10^{3} \, \mathrm{eV}^2
$
is presented in Fig.\ref{fig3}
(the solid and dashed lines, respectively).
The thick solid line in Fig.\ref{fig3}
represents the unitarity upper bound
$
|\langle{m}\rangle|
\leq
\sqrt{ \Delta{m}^{2} }
$.

As can be seen from Fig.\ref{fig3},
the upper bound for the effective Majorana
mass $|\langle{m}\rangle|$
depends rather strongly on the value of
$\Delta{m}^{2}$
(whose square root is equal to
the heaviest mass $m_3$).
From Fig.\ref{fig3} one can see that
if $\Delta{m}^{2}$ is less than
$ 10 \, \mathrm{eV}^2 $,
the effective Majorana mass
$|\langle{m}\rangle|$
is smaller than
$ 10^{-1} \, \mathrm{eV} $.
Figure \ref{fig3}
show also that
if $\Delta{m}^{2}$ is less than
$ 2 \, \mathrm{eV}^2 $,
from exclusion plots of
the Bugey and CHOOZ experiments
it follows that
$ |\langle{m}\rangle| \lesssim 3 \times 10^{-2} \, \mathrm{eV} $.
 
Up to now we have considered only
the data of reactor neutrino experiments.
Let us now take into account
also the results of the LSND experiment \cite{LSND96}.
The data of this experiment fix an allowed region of $\Delta{m}^{2}$.
Combined with
the negative results of
the Bugey \cite{Bugey95} and BNL E776 \cite{BNLE776} experiments,
the allowed plot of 
the LSND experiment imply that 
$\Delta{m}^{2}$ lies in the range
\begin{equation}
0.3 \, \mathrm{eV}^2
\lesssim
\Delta{m}^2
\lesssim
2.2 \, \mathrm{eV}^2
\,.
\label{LSNDrange}
\end{equation}
The corresponding region of allowed values of
$|\langle{m}\rangle|$
is represented by the shadowed region
in Fig.\ref{fig3}.
One can see that the results of the LSND experiment,
together with the negative results of other SBL experiments,
imply that the value of $|\langle{m}\rangle|$ is very small:
$ |\langle{m}\rangle| \lesssim 3 \times 10^{-2} \, \mathrm{eV} $.

Thus, we conclude that
if massive neutrinos are Majorana particles
and
if there is a hierarchy of neutrino masses,
the existing data of reactor neutrino experiments 
imply a strong constraint on the parameter
$|\langle{m}\rangle|$:
$ |\langle{m}\rangle| \lesssim 10^{-1} \, \mathrm{eV} $
for
$ \Delta{m}^{2} \lesssim 10 \, \mathrm{eV}^2 $.
Let us stress that the value
$ |\langle{m}\rangle| \sim 10^{-1} \, \mathrm{eV} $
corresponds to the sensitivity 
of the next generation of $(\beta\beta)_{0\nu}$ decay experiments
\cite{Heidelberg-Moscow,futureBB}.

If the results of the LSND experiment
are confirmed by future experiments,
the upper bound for the parameter
$|\langle{m}\rangle|$
is about
$ 3 \times 10^{-2} \, \mathrm{eV} $.
Such small values of
$|\langle{m}\rangle|$
can be explored only by
$(\beta\beta)_{0\nu}$ decay
experiments of future generations
(see Ref.\cite{Klapdor-Erice}).

\section{Constraints from atmospheric neutrino experiments}
\label{Constraints from atmospheric neutrino experiments}

In the previous Section we have obtained
constraints on the Majorana parameter
$|\langle{m}\rangle|$
from the results of reactor experiments and
of the LSND experiment.
In this Section
we present the allowed region
for the parameter
$|\langle{m}\rangle|$
obtained from the data of atmospheric neutrino experiments
in the scheme with mixing of three Majorana neutrinos and
a neutrino mass hierarchy.
The ratio of muon and electron atmospheric neutrino events
has been found to be significantly smaller than the expected one
in the
Kamiokande \cite{Kam-atm},
IMB \cite{IMB}
and
Soudan \cite{Soudan}
experiments.
For
the double ratio
$
R = (\mu/e)_{\mathrm{data}}/(\mu/e)_{\mathrm{MC}}
$
($(\mu/e)_{\mathrm{MC}}$
is the Monte-Carlo
calculated ratio of muon and electron events
under the assumption that neutrinos do not oscillate),
in the regions of neutrino energies less than 1.3 GeV (sub-GeV)
and more than 1.3 GeV (multi-GeV)
the Kamiokande collaboration found
\begin{equation}
R_{\mathrm{Kamiokande}}^{\mathrm{sub-GeV}}
=
0.60 \mbox{}^{+0.06}_{-0.05} \pm 0.05
\,,
\qquad
\qquad
R_{\mathrm{Kamiokande}}^{\mathrm{multi-GeV}}
=
0.57 \mbox{}^{+0.08}_{-0.07} \pm 0.07
\,.
\label{27}
\end{equation}
The
IMB \cite{IMB}
and
Soudan \cite{Soudan}
collaborations found
\begin{equation}
R_{\mathrm{IMB}}
=
0.54 \pm 0.05 \pm 0.12
\,,
\qquad
\qquad
R_{\mathrm{Soudan}}
=
0.75 \pm 0.16 \pm 0.10
\,.
\label{29}
\end{equation}
On the other hand,
the values of the double ratio found
in the Frejus \cite{Frejus} and NUSEX \cite{NUSEX}
experiments,
\begin{equation}
R_{\mathrm{Frejus}}
=
0.99 \pm 0.13 \pm 0.08
\,,
\qquad
\qquad
R_{\mathrm{NUSEX}}
=
1.04 \pm 0.25
\,,
\label{31}
\end{equation}
are compatible with unity
(but cannot exclude the atmospheric neutrino anomaly
because of large errors).

The existence of the atmospheric neutrino anomaly was recently confirmed
by the results of the
high statistics Super-Kamiokande experiment
\cite{SK-atm}:
\begin{equation}
R_{\mathrm{Super-K}}^{\mathrm{sub-GeV}}
=
0.635 \mbox{}^{+0.034}_{-0.033} \pm 0.010 \pm 0.052
\,,
\quad
R_{\mathrm{Super-K}}^{\mathrm{multi-GeV}}
=
0.604 \mbox{}^{+0.065}_{-0.058} \pm 0.018 \pm 0.065
\,.
\label{32}
\end{equation}
Here the three errors are,
respectively,
the statistical errors of the data,
the statistical error of the Monte Carlo
and the systematic error.

The results of atmospheric neutrino experiments
can be explained by neutrino oscillations.
The recent results of the CHOOZ experiment \cite{CHOOZ97}
exclude the possibility of
$\nu_\mu\leftrightarrows\nu_e$
oscillations.
In the framework of two flavor
$\nu_\mu\to\nu_\tau$
oscillations,
the following 90\% CL allowed ranges for the oscillation parameters
were found by the analysis of
the Kamiokande data \cite{Kam-atm}:
\begin{equation}
5 \times 10^{-3} \, \mbox{eV}^2
\lesssim
\Delta{m}^2
\lesssim
3 \times 10^{-2} \, \mbox{eV}^2
\,,
\qquad
0.7
\lesssim
\sin^2 2\vartheta
\lesssim
1
\,.
\label{33}
\end{equation}
The preliminary analysis of the Super-Kamiokande data \cite{SK-atm}
indicate
the following 90\% CL allowed ranges for the oscillation parameters:
\begin{equation}
3 \times 10^{-4} \, \mbox{eV}^2
\lesssim
\Delta{m}^2
\lesssim
6 \times 10^{-3} \, \mbox{eV}^2
\,,
\qquad
0.8
\lesssim
\sin^2 2\vartheta
\lesssim
1
\,.
\label{34}
\end{equation}
The values of
$\Delta{m}^2$
allowed by the Super-Kamiokande data
are significantly smaller than those
allowed by the Kamiokande data.
However,
the two allowed ranges of
$\Delta{m}^2$
overlap at
$ \Delta{m}^2 \simeq 5 \times 10^{-3} \, \mbox{eV}^2 $,
indicating that the two experimental results are compatible.

In Section
\ref{Constraints from reactor neutrino experiments and the LSND experiment}
we obtained restrictions on the parameter
$|\langle{m}\rangle|$
from the exclusion plots of reactor experiments
and from the allowed plot of the LSND experiment.
Here we present the
allowed region of the Majorana parameter
$|\langle{m}\rangle|$
obtained from the results of a $\chi^2$ analysis of
the Kamiokande atmospheric neutrino
data
in the model
with mixing of three neutrinos and a neutrino mass hierarchy
\cite{GKM97}.
In this case,
the oscillation probabilities of atmospheric neutrinos
depend on three parameters:
$\Delta{m}^{2}$,
$|U_{e3}|^2$
and
$|U_{\mu3}|^2$
($|U_{\tau3}|^2=1-|U_{\mu3}|^2-|U_{\mu3}|^2$).
The matter effect
for the atmospheric neutrinos
reaching the Kamiokande detector from below
has been taken into account.
The presence of matter
is important because it modifies the phases
of neutrino oscillations
\cite{Pantaleone94}
and its effect is to enlarge
the allowed region towards low values of
$\Delta{m}^{2}$
(see Ref.\cite{GKM97}).
The best fit of the Kamiokande data
is obtained for
$ \Delta{m}^{2} = 2.5 \times 10^{-2} \, \mathrm{eV}^2 $,
$ |U_{e3}|^2 = 0.26 $
and
$ |U_{\mu3}|^2 = 0.49 $,
with
$ \chi^2 = 6.9 $
for 9 degrees of freedom,
corresponding to a CL of 65\%.

The range \emph{allowed} at 90\% CL
in the
$|\langle{m}\rangle|$--$\Delta{m}^{2}$
plane is shown in Fig.\ref{fig4}
as the vertically shadowed region.
The solid and dashed lines in Fig.\ref{fig4}
represent the upper bounds
obtained with Eq.(\ref{26}) from the
90\% CL exclusion plots
of the
Bugey \cite{Bugey95}
and
CHOOZ \cite{CHOOZ97}
experiments,
respectively.
The thick solid line
represents the unitarity upper bound
$
|\langle{m}\rangle|
\leq
\sqrt{ \Delta{m}^{2} }
$.
From Fig.\ref{fig4}
it can be seen that
the results of the Kamiokande experiment,
together with the exclusion plots of the Bugey and CHOOZ experiments,
imply that
\begin{equation}
|\langle{m}\rangle|
\lesssim
8 \times 10^{-3} \, \mathrm{eV}
\;.
\label{35}
\end{equation}

The horizontally shadowed region in Fig.\ref{fig4}
indicates the range (\ref{34}) of
$\Delta{m}^{2}$
allowed at 90\% CL by the preliminary analysis
of the data of the Super-Kamiokande experiment
\cite{SK-atm}.
This range covers values of $\Delta{m}^{2}$
smaller by an order of magnitude
with respect to the
range of $\Delta{m}^{2}$
allowed by the Kamiokande data.
However,
the two allowed ranges have an overlap
around
$ \Delta{m}^{2} \simeq 5 \times 10^{-3} \, \mathrm{eV}^2 $.
If this is the value of
$\Delta{m}^{2}$,
the exclusion curve of the CHOOZ experiment puts a
very strong constraint on
$|\langle{m}\rangle|$:
\begin{equation}
|\langle{m}\rangle|
\lesssim
3 \times 10^{-3} \, \mathrm{eV}
\;.
\label{36}
\end{equation}

Thus we can conclude that
in all possible
scenarios with mixing of three massive Majorana neutrinos
and a neutrino mass hierarchy,
the existing neutrino oscillation data
imply that the effective Majorana mass,
which characterizes the matrix
element of $(\beta\beta)_{0\nu}$ decays,
is very small.

\section{Non-hierarchical neutrino mass spectra}
\label{Non-hierarchical neutrino mass spectra}

In this Section we consider the following two possibilities:

I. \emph{``Inverted'' mass hierarchy of three neutrinos.}
In the previous Sections 
we have assumed that there are
three Majorana neutrinos
with a hierarchy of masses
and that $\Delta{m}^2_{21}$ is relevant for the suppression
of the flux of solar $\nu_e$'s.
Another possibility to explain
the solar neutrino data
in the framework of three neutrino mixing
is to assume that the neutrino mass spectrum has the form
\cite{three-inverted,BGKP}
\begin{equation}
m_1 \ll m_2 \simeq m_3
\,,
\label{41}
\end{equation}
and $\Delta{m}^2_{32}$
is relevant for the suppression of the flux of solar $\nu_e$'s.
In this case,
SBL neutrino oscillations are described
by the expressions (\ref{17}) and (\ref{19}),
with the change
$ U_{\ell3} \to U_{\ell1} $.
From the exclusion plots of reactor experiments
it follows that
\begin{equation}
|U_{e1}|^2 \leq a_e^0
\,,
\label{42}
\end{equation}
with $a_e^0$ given by Eq.(\ref{24}).
The value of $a_e^0$ depends on
$ \Delta{m}^2 \equiv m_3^2 - m_1^2 \simeq m_3^2 $.
From Fig.\ref{fig1}
one can see that
$a_e^0$
is small for
$\Delta{m}^{2}$
in the range (\ref{wide}).
In this case,
the effective Majorana mass is given by
\begin{equation}
|\langle{m}\rangle|
\simeq
| U_{e2}^2 + U_{e3}^2 | \, \sqrt{\Delta{m}^2}
\,.
\label{43}
\end{equation}
If CP is conserved in the lepton sector and
the relative CP parity of $\nu_2$ and $\nu_3$ is equal to unity,
$|\langle{m}\rangle|$
is (practically) equal to $\sqrt{\Delta{m}^2}$
\cite{PS94,BGKP}.
In general, we have
\begin{equation}
|\langle{m}\rangle|
\leq
\sqrt{\Delta{m}^2}
\,.
\label{44}
\end{equation}
Thus, in the case of the neutrino mass spectrum (\ref{41}),
the upper bound for the effective Majorana mass can be in the eV region. 
If the spectrum (\ref{41}) is realized in nature,
from the inequality (\ref{42}) it follows
also that neutrino mass $m(^3\mathrm{H})$
measured in $^3\mathrm{H}$-decay experiments
is practically equal to the heaviest mass \cite{BGKP}:
\begin{equation}
m(^3\mathrm{H})
\simeq
\sqrt{\Delta{m}^2}
\,.
\label{45}
\end{equation}
Let us notice that in the case of a hierarchy of three neutrino masses
the contribution of the term that depends on
$ m_3 \simeq \sqrt{\Delta{m}^2} $
to the $\beta$-spectrum of $^3\mathrm{H}$
is suppressed by the factor
$ |U_{e3}|^2 \leq a_e^0 $.
Therefore,
the observation of the effect of a neutrino mass in
the experiments measuring
the high-energy part of the $\beta$-spectrum of $^3\mathrm{H}$
\cite{troitsk,mainz}
would be an indication in favor of
the neutrino spectrum 
(\ref{41})
with an ``inverted'' mass hierarchy.

II. \emph{Four massive neutrinos.}
All the existing indications in favor of neutrino mixing 
(solar neutrinos, atmospheric neutrinos, LSND)
cannot be described by any scheme with three massive 
neutrinos
\cite{BGG96,OY96,FLMS97,Erice}.
If we take all data seriously,
we need to consider schemes
of mixing with (at least) four massive neutrinos
\cite{four},
that include
not only
$\nu_e$,
$\nu_\mu$
and
$\nu_\tau$,
but also (at least) one sterile neutrino.
In \cite{BGG96,OY96}
it was shown that among all the
possible mass spectra of four neutrinos only two can accommodate all the
existing data:
\begin{equation}
(\mathrm{A})
\qquad
\underbrace{
\overbrace{m_1 < m_2}^{\mathrm{atm}}
\ll
\overbrace{m_3 < m_4}^{\mathrm{sun}}
}_{\mathrm{LSND}}
\,,
\qquad \qquad
(\mathrm{B})
\qquad
\underbrace{
\overbrace{m_1 < m_2}^{\mathrm{sun}}
\ll
\overbrace{m_3 < m_4}^{\mathrm{atm}}
}_{\mathrm{LSND}}
\,.
\label{46}
\end{equation}
In the case of scheme A,
$\Delta{m}^2_{21}$
is relevant for the atmospheric neutrino
anomaly and
$\Delta{m}^2_{43}$
for the suppression of solar $\nu_e$'s,
whereas in the scheme B
the roles of
$\Delta{m}^2_{21}$
and
$\Delta{m}^2_{43}$
are reversed.
In both schemes two groups of close 
masses are separated by the "LSND gap"
of the order of 1 eV.
In the scheme B,
the upper bound
for the effective Majorana mass is given by
\begin{equation}
|\langle{m}\rangle|
\leq
a_e^0 \, \sqrt{\Delta{m}^2}
\,,
\label{47}
\end{equation}
with
$ \Delta{m}^2 \equiv m_4^2 - m_1^2 \simeq m_4^2 $.
Hence,
in the scheme B
the effective Majorana mass
$|\langle{m}\rangle|$
must satisfy the constraints discussed in
Section \ref{Constraints from reactor neutrino experiments and the LSND experiment}
and presented in Fig.\ref{fig3}.
This means that
in the scheme B
the contribution of Majorana neutrino masses
to the amplitude of $(\beta\beta)_{0\nu}$-decay
is strongly suppressed.
In the scheme A,
the effective Majorana mass is bounded by
\begin{equation}
|\langle{m}\rangle|
\leq
\sum_{i=3,4}
|U_{ei}|^2 \, \sqrt{\Delta{m}^2}
\leq
\sqrt{\Delta{m}^2}
\,.
\label{48}
\end{equation}
Hence, no a priori
suppression of the Majorana mass contribution to $(\beta\beta)_{0\nu}$-decay
is expected in the scheme A.

Also the values of the effective neutrino mass
$m(^3\mathrm{H})$
measured in the experiments that investigate the high-energy part
of the $\beta$-spectrum of $^3\mathrm{H}$
\cite{troitsk,mainz}
are different in the schemes A and B.
In the scheme A we have
$ m(^3\mathrm{H}) \simeq m_4 \simeq \sqrt{\Delta{m}^2} $,
whereas
in the scheme B
the contribution of the term that depends on the heavy masses
$ m_3 \simeq m_4 $
to the $\beta$-spectrum of $^3\mathrm{H}$
is suppressed by the factor
$ \sum_{i=3,4} |U_{ei}|^2 \leq a_e^0 $.

\section{Conclusions}
\label{Conclusions}

We have obtained various constraints on the parameter
$|\langle{m}\rangle|$
(that characterizes the contribution of Majorana
neutrino masses to the matrix element of neutrinoless double-beta decay)
from the results of neutrino oscillation experiments.
We have shown that
in the scheme with mixing of three
Majorana neutrinos and a mass hierarchy
(which corresponds to the
see-saw mechanism for the generation of neutrino masses)
the results of neutrino oscillation experiments
put rather severe restrictions on the value of
$|\langle{m}\rangle|$.
The numerical value of the upper bound for
$|\langle{m}\rangle|$
depends rather strongly
on the value of the parameter
$ \Delta{m}^2 \equiv m_3^2 - m_1^2 $.
If we take into account only the results of
SBL reactor experiments
and the results of solar neutrino experiments,
we can conclude that
$ |\langle{m}\rangle| \lesssim 10^{-1} \, \mathrm{eV} $
for
$ \Delta{m}^2 \lesssim 10 \, \mathrm{eV}^2 $.
From the new results of the first LBL experiment CHOOZ
and from the exclusion curve of the Bugey experiment
it follows that
for
$ \Delta{m}^2 \lesssim 2 \, \mathrm{eV}^2 $
the parameter
$|\langle{m}\rangle|$
is less than
$ 3 \times 10^{-2} \, \mathrm{eV} $. 
If we take into account the results of the LSND experiment,
we come to the conclusion that
$ |\langle{m}\rangle| \lesssim 3 \times 10^{-2} \, \mathrm{eV} $.

We have also calculated the region of the parameter
$|\langle{m}\rangle|$
allowed by the data of the Kamiokande atmospheric neutrino experiment,
using the recent three-neutrino $\chi^2$ analysis
presented in Ref.\cite{GKM97}.
Taking into account this allowed range of
$|\langle{m}\rangle|$
and the constraints
obtained from the results of the Bugey and CHOOZ experiments,
we conclude that 
very small values of
$|\langle{m}\rangle|$
are allowed:
$ |\langle{m}\rangle| \lesssim 8 \times 10^{-3} \, \mathrm{eV} $.
Taking into account also
the results of the preliminary analysis of Super-Kamiokande data,
an even stronger constraint can be placed:
$ |\langle{m}\rangle| \lesssim 3 \times 10^{-3} \, \mathrm{eV} $.

The constraints on the value of the effective Majorana mass
$|\langle{m}\rangle|$
that follow from the results of neutrino oscillation experiments
must be taken into account
in the interpretation of the data of
$(\beta\beta)_{0\nu}$-decay experiments.
The observation of neutrinoless double-beta decay
with a probability that corresponds to
$ |\langle{m}\rangle| \gtrsim 10^{-1} \, \mathrm{eV} $
(which is the sensitivity of future
$(\beta\beta)_{0\nu}$-decay experiments)
would imply that the spectrum of three neutrinos does not follow
a hierarchical pattern and the neutrino masses are not of see-saw origin,
or that
there are more than three massive neutrinos.
This observation could also
imply that non-standard mechanisms for the violation of lepton 
number, 
such as
right-handed currents (see Refs.\cite{Mohapatra-Pal,Mohapatra95,HKP}),
supersymmetry with violation of R-parity \cite{BM95,Mohapatra95,HKK-susy,FKSS},
and others \cite{HKK-others,PCSW},
are responsible for neutrinoless double-beta decay.

Thus,
the observation of $(\beta\beta)_{0\nu}$-decay could allow us to obtain 
information 
not only about the nature of massive neutrinos (Dirac or Majorana?),
but also about the pattern of the mass spectrum of neutrinos
and/or
about non-standard mechanisms of violation of the lepton number.

\acknowledgments

S.M.B. would like to acknowledge supports from
the Dyson Visiting
Professor Funds at the Institute for Advanced Study.

\begin{figure}[h]
\protect\caption{The $\Delta{m}^2$
versus $a_{e}^{0}$
plot
obtained from
the 90\% CL exclusion plots of the
Bugey \protect\cite{Bugey95}
and
CHOOZ \protect\cite{CHOOZ97}
reactor neutrino oscillation experiments
(see Eq.(\ref{24})).}
\label{fig1}
\end{figure}

\begin{figure}[h]
\protect\caption{The lower bound
$ (P_{\nu_e\to\nu_e}^{\mathrm{sun}})_{\mathrm{min}} $
for the probability
of solar $\nu_e$'s to survive in the 
case of a large value of the parameter
$|U_{e3}|^2$
($ \geq 1 - a_{e}^{0} $).
The values of
$a_{e}^{0}$
are obtained from
the 90\% CL exclusion plots of the
Bugey \protect\cite{Bugey95}
and
CHOOZ \protect\cite{CHOOZ97}
reactor neutrino oscillation experiments.}
\label{fig2}
\end{figure}

\begin{figure}[h]
\protect\caption{Upper bounds for
the effective Majorana mass
$|\langle{m}\rangle|$
obtained from the 90\% CL exclusion plots of
the Bugey (solid line)
and CHOOZ (dashed line)
neutrino reactor experiments.
The shadowed region
corresponds to the range of $\Delta{m}^2$
allowed at 90\% CL by the results of the LSND experiment,
taking into account the results of
all the other SBL experiments.
The thick solid line
represents the unitarity upper bound
$
|\langle{m}\rangle|
\leq
\sqrt{ \Delta{m}^{2} }
$.}
\label{fig3}
\end{figure}

\begin{figure}[h]
\protect\caption{Upper bounds for
the effective Majorana mass
$|\langle{m}\rangle|$
obtained from the 90\% CL exclusion plots of
the Bugey (solid line)
and CHOOZ (dashed line)
neutrino reactor experiments.
The vertically shadowed region is allowed at 90\% CL
by the data of
the Kamiokande atmospheric neutrino experiment.
The horizontally shadowed region
corresponds to the range of $\Delta{m}^2$
allowed at 90\% CL by the preliminary analysis
of the data of the Super-Kamiokande experiment.
The thick solid line
represents the unitarity upper bound
$
|\langle{m}\rangle|
\leq
\sqrt{ \Delta{m}^{2} }
$.}
\label{fig4}
\end{figure}


\begin{minipage}{0.95\textwidth}
\begin{center}
\includegraphics[bb=30 30 550 780,width=0.95\textwidth]{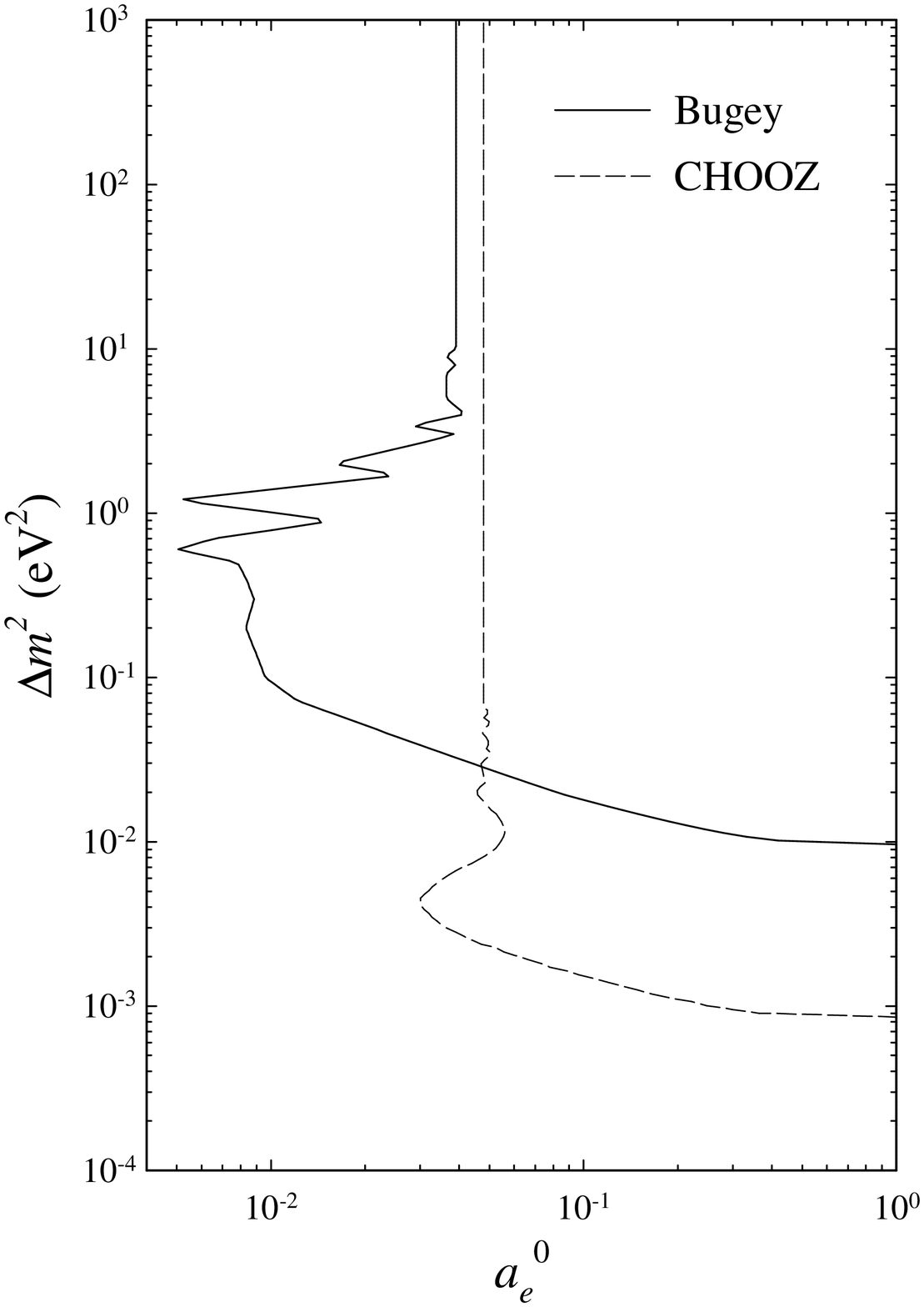}
\end{center}
\end{minipage}
\begin{center}
\Large Figure~\ref{fig1}
\end{center}

\begin{minipage}{0.95\textwidth}
\begin{center}
\includegraphics[bb=30 30 550 780,width=0.95\textwidth]{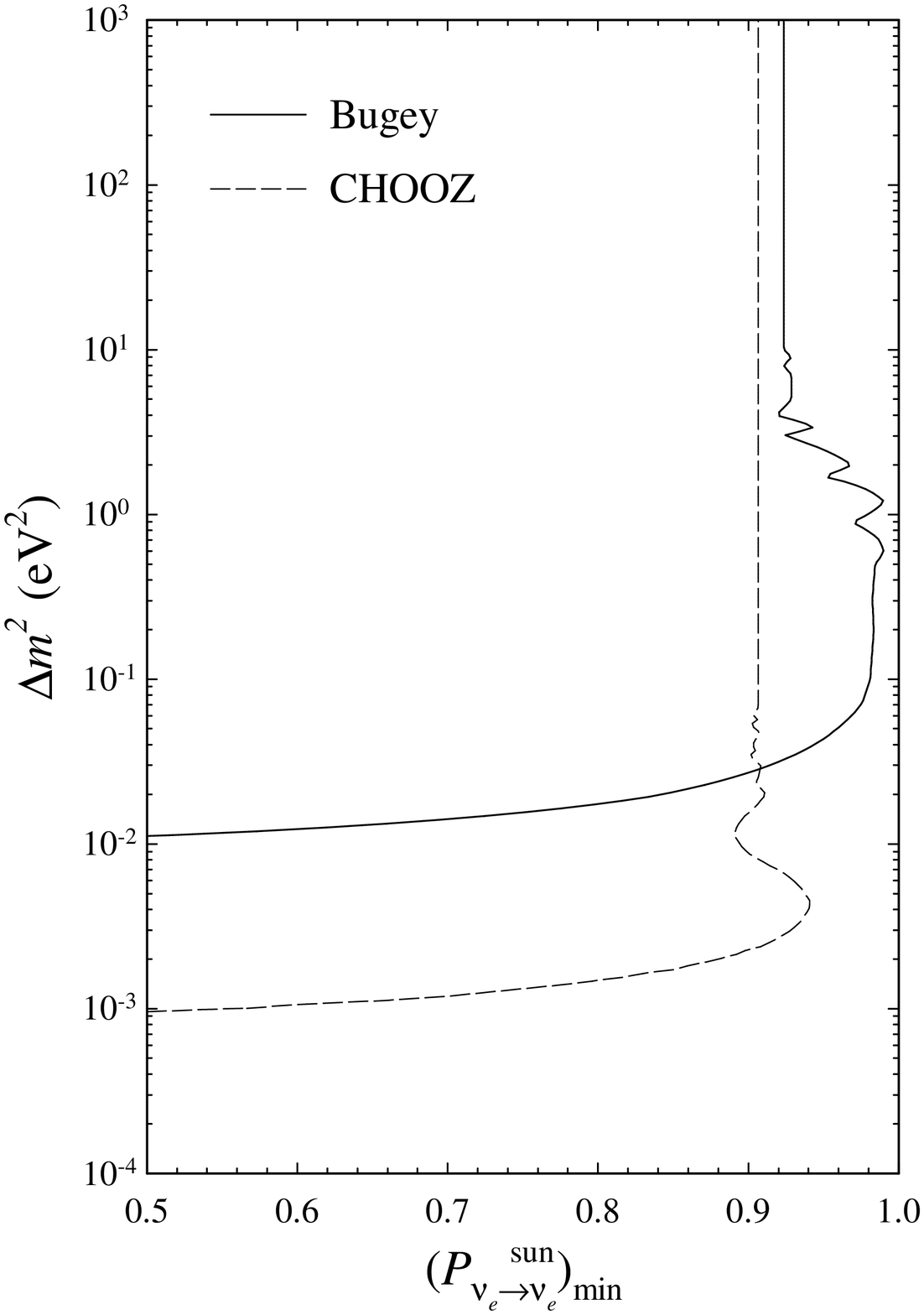}
\end{center}
\end{minipage}
\begin{center}
\Large Figure~\ref{fig2}
\end{center}

\begin{minipage}{0.95\textwidth}
\begin{center}
\includegraphics[bb=30 30 550 780,width=0.95\textwidth]{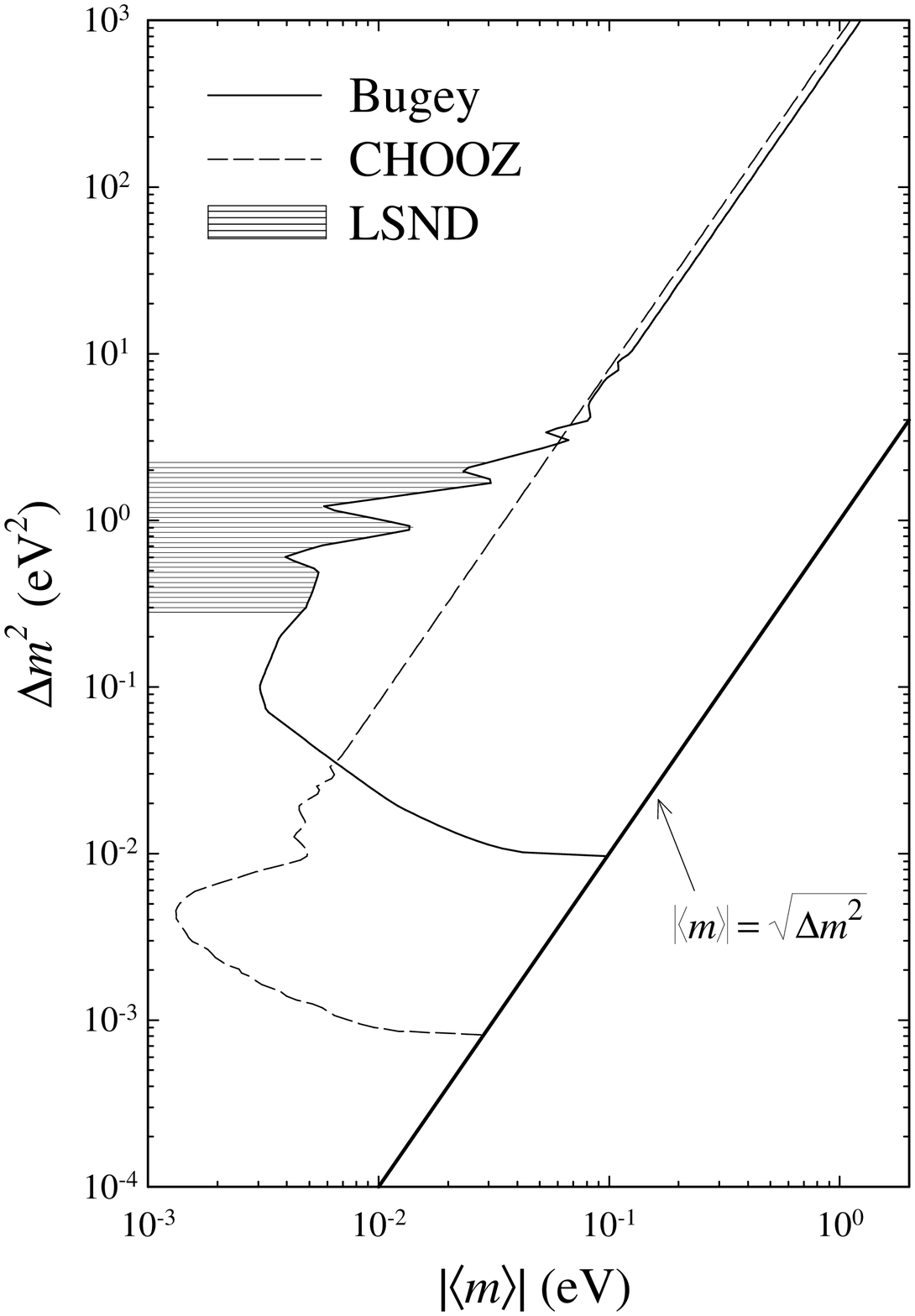}
\end{center}
\end{minipage}
\begin{center}
\Large Figure~\ref{fig3}
\end{center}

\begin{minipage}{0.95\textwidth}
\begin{center}
\includegraphics[bb=30 30 550 780,width=0.95\textwidth]{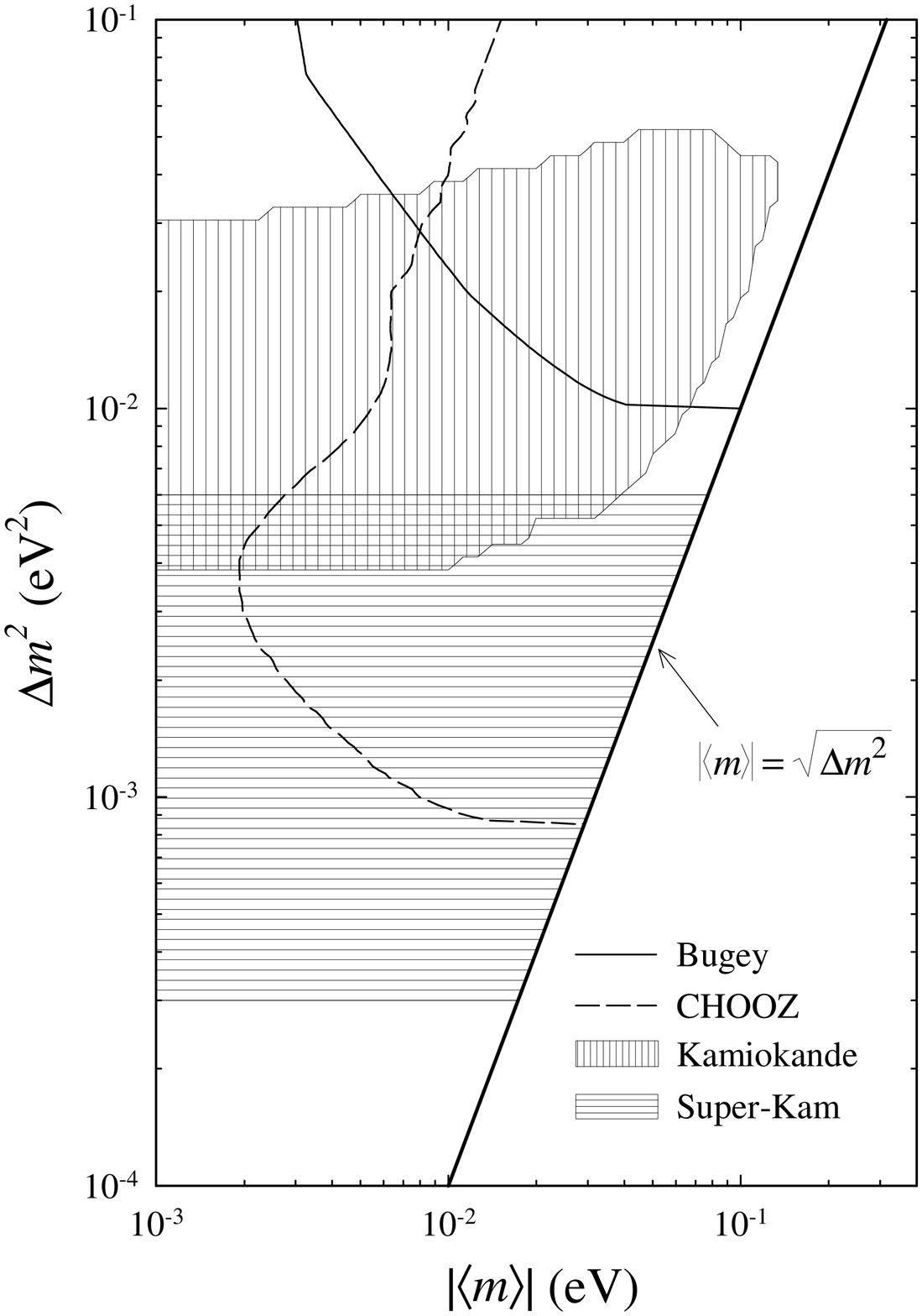}
\end{center}
\end{minipage}
\begin{center}
\Large Figure~\ref{fig4}
\end{center}

\end{document}